\documentclass[twocolumn]{aastex701}
\usepackage{natbib}
\usepackage{amsmath}

\bibpunct{(}{)}{;}{a}{}{,}

\newcommand{\be}{\begin{eqnarray}}
\newcommand{\ee}{\end{eqnarray}}

\begin{document}

\title{A Multi-Wavelength Search for Nearby Black Holes Accreting from the Interstellar Medium}

\author[0009-0009-0287-2535]{Abdurakhmon Nosirov}
\affiliation{Center for Astronomy and Astrophysics, Department of Physics, Fudan University, Shanghai 200438, China}
\affiliation{Institut f\"ur Astronomie und Astrophysik, Eberhard-Karls Universit\"at T\"ubingen, D-72076 T\"ubingen, Germany}
\email{abdurakhmonn24@m.fudan.edu.cn}  

\author[0000-0002-3180-9502]{Cosimo Bambi}
\altaffiliation{bambi@fudan.edu.cn}
\affiliation{Center for Astronomy and Astrophysics, Department of Physics, Fudan University, Shanghai 200438, China}
\affiliation{Institute for Advanced Studies, New Uzbekistan University, Tashkent 100007, Uzbekistan}
\email{bambi@fudan.edu.cn} 

\author[0000-0003-4187-9560]{Andrea Santangelo}
\affiliation{Institut f\"ur Astronomie und Astrophysik, Eberhard-Karls Universit\"at T\"ubingen, D-72076 T\"ubingen, Germany}
\email{andrea.santangelo@uni-tuebingen.de}

\author[0009-0000-9251-1649]{Yi Fan}
\affiliation{Center for Astronomy and Astrophysics, Department of Physics, Fudan University, Shanghai 200438, China}
\email{yfan26@m.fudan.edu.cn} 

\author[0000-0003-3319-166X]{Leda Gao}
\affiliation{Center for Astronomy and Astrophysics, Department of Physics, Fudan University, Shanghai 200438, China}
\email{ledagao@fudan.edu.cn} 

\author[0000-0001-6459-8599]{Jos de Bruijne}
\affiliation{European Space Agency (ESA), European Space Research and Technology Centre (ESTEC), NL-2201 AZ Noordwijk, The Netherlands}
\email{Jos.De.Bruijne@esa.int}

\author[0009-0009-2549-1161]{Ningyue Fan}
\affiliation{Department of Physics, Stanford University, Stanford, CA 94305, USA}
\affiliation{Kavli Institute for Particle Astrophysics and Cosmology, Stanford University, Stanford, CA 94305, USA}
\email{nyfan@stanford.edu}

\author[0000-0002-9639-4352]{Jiachen Jiang}
\affiliation{Department of Physics, University of Warwick, Coventry CV4 7AL, UK}
\email{jcjiang12@outlook.com}

\author[0000-0001-7519-1700]{Federico Marocco}
\affiliation{IPAC, California Institute of Technology, Pasadena, CA 91125, USA}
\email{federico@ipac.caltech.edu}

\author[0000-0001-9969-2091]{Fu-Guo Xie}
\affiliation{Shanghai Astronomical Observatory, Chinese Academy of Sciences, Shanghai 200030, China}
\email{fgxie@shao.ac.cn}

\begin{abstract}
Theoretical models suggest that the Galaxy hosts a large population of stellar-mass black holes, and most of these objects are isolated, without any companion. Within 15~pc of the Solar System, there may be a few black holes. Here we report our attempt to find nearby black hole candidates accreting from the interstellar medium in existing catalogs. We start by selecting high proper motion sources in the IR catalog CatWISE2020, and then we use the optical catalog NSC~DR2, the IR catalogs AllWISE, VHS, and UHS, and the radio catalogs VLASS~QL1, RACS-mid, and RACS-high. Eventually, we find 13~``gold'' candidates, 188~``silver'' candidates, and 2,186~``bronze'' candidates. The nature of these sources can be confirmed only by follow-up observations.
\end{abstract}


\section{Introduction}

Stellar-mass black holes are the final product of the evolution of stars with masses $M_{\rm star} \gtrsim 20~M_\odot$~\citep{1999ApJ...522..413F,2002RvMP...74.1015W}. Population synthesis studies suggest that in the Galaxy there may be $10^8$-$10^9$~stellar-mass black holes formed from the gravitational collapse of massive stars~\citep{1996ApJ...457..834T,2020A&A...638A..94O,2026arXiv260722814W,2026arXiv260929211H,2026arXiv261007152K} and that most of these objects are isolated, without any companion. Indeed, while about 70\% of the massive stars are in binary systems, the binary survival probability is low. In the case of compact binaries, the two stars of the system can likely merge during the red supergiant phase of the massive component. In the case of wide binaries, the supernova explosion of the massive star can likely disrupt the system.

\citet{2020A&A...638A..94O} predict that $\sim$92\% of the black holes in the Galactic disk are isolated, $\sim$6.4\% are in a binary system with another black hole, $\sim$1.6\% are in a binary system with a neutron star or white dwarf, and only $\sim$0.16\% of the black holes are in a binary system with a normal star. Today, we know of fewer than 80~stellar-mass black holes in the Galaxy and the vast majority of them are in X-ray binary systems \citep[for a review, see, e.g.,][and references therein]{2025Symm...17.1393B}: while these systems are extremely rare and represent only a small fraction of the binaries black hole--normal star, they are much easier to identify.

Within 15~pc (about 50~ly) of the Solar System, there may be a few stellar-mass black holes~\citep{2013MNRAS.430.1538F,2025ApJ...988L..12M,2026ApJ..1004...21N}. For a rough estimate, we can assume that the Galactic disk is a cylinder with radius $R = 10$~kpc and height $H = 0.5$~kpc: the volume of such a cylinder is $157$~kpc$^3$ and includes 80-90\% of the stars of the Galactic disk. \citet{2020A&A...638A..94O} predict $N = 1.0 \cdot 10^8$~black holes in the Galactic disk; the mean black hole density is thus $N/V = 1/1570$~pc$^{-3}$. The sphere with radius 15~pc centered on the Solar System has a volume of 14130~pc$^3$ and we can expect $\sim$14130/1570 = 9~black holes.

This simple calculation may somewhat overestimate the local black hole density because $i)$ the stellar density is not constant over the Galactic disk, and the Solar System is at 8~kpc from the Galactic Center, where the local stellar density (and consequently the local black hole density) is lower than the mean one; $ii)$ if black holes receive a kick during the supernova explosion, they are spread over a thicker disk than the stellar disk. \citet{2026arXiv260722814W} predict a local black hole density of $1.6 \cdot 10^{-4}$~pc$^{-3}$ in their fiducial model, and this corresponds to 2-3~black holes within 15~pc of the Solar System. \citet{2026arXiv261007152K} predict more than twice as many black holes as \citet{2026arXiv260722814W}, increasing the local number of black holes by the same factor.

Isolated black holes can potentially be detected from the radiation emitted by their accretion from the interstellar medium~\citep{1971SvA....15..377S,1975A&A....44...59M,1985MNRAS.217...77M,1998ApJ...495L..85F,2005MNRAS.360L..30M,2013MNRAS.430.1538F,2018MNRAS.477..791T,2021MNRAS.505.4036S,2025ApJ...988L..12M,2025A&A...700A..49M,2026ApJ..1004...21N}. \citet{2025ApJ...988L..12M} pointed out that the identification of one of these black holes may be very challenging: the object may just appear as a faint source on the sky with a relatively featureless spectrum. Identification will necessarily require observations covering the radio, IR, optical, and X-ray bands in order to rule out alternative interpretations and confirm the black hole nature of the source.

Within 15~pc of the Solar System, 80-95\% of the volume is filled by a hot, low-density interstellar medium, and the remaining 5-20\% is occupied by a complex of 15 warm, partially ionized clouds~\citep{2008ApJ...673..283R}. If an isolated black hole were located inside one of these clouds, it could produce detectable electromagnetic emission through accretion from the interstellar medium; outside these clouds, however, the accretion rate would likely be too low to generate a signal detectable by current or near-future observational facilities~\citep{2026ApJ..1004...21N}. \citet{2026ApJ..1004...21N} also looked for potential candidate black holes within 15~pc of the Solar System starting from the Gaia~DR3 catalog~\citep{2023A&A...674A...1G}, without finding any promising source. The specific interest in finding a black hole within 15~pc of the Solar System is motivated by the future possibility of an interstellar mission to such a black hole, with the goal of testing black holes and General Relativity at a level likely impossible with observational facilities on the Earth or in the Solar System~\citep{2025iSci...28k3142B,2026arXiv260710982B,2026EPJC...86.1192G,2026EPJC...86.1197F}.

Here we present a new study devoted to finding nearby black hole candidates accreting from the interstellar medium, starting from what we learnt from \citet{2026ApJ..1004...21N}. We start from the IR catalog CatWISE2020~\citep{2020ApJS..247...69E,2021ApJS..253....8M} and we select high proper motion sources, without requiring any distance estimate, because faint sources often lack reliable parallax measurements. We then cross-match our sources with the optical catalog NSC~DR2~\citep{2021AJ....161..192N}, the IR catalogs AllWISE~\citep{2013wise.rept....1C}, VHS~\citep{2013Msngr.154...35M}, and UHS~\citep{2018MNRAS.473.5113D,2025AJ....170...86S}, and the radio catalogs VLASS~QL1~\citep{2020PASP..132c5001L}, RACS-mid, and RACS-high~\citep{2024PASA...41....3D,2025PASA...42...38D}. We fit the flux measurements of every source with the public model {\tt LLAGNSED}~\citep{2021ApJ...923..260P}, which is designed to calculate spectra of black holes with a very low mass accretion rate within an advection dominated accretion flow (ADAF) model. Eventually, we find 13~``gold'' candidates, 188~``silver'' candidates, and 2,186~``bronze'' candidates. The nature of these sources cannot be confirmed without follow-up observations.

Our paper is organized as follows. In Section~\ref{s-spectra}, we review the theoretical predictions for the spectra expected from isolated black holes accreting from the interstellar medium. In Section~\ref{s-search}, we describe our multi-wavelength strategy to look for nearby isolated black holes accreting from the interstellar medium. Discussion and conclusions are reported in Section~\ref{s-d-c}.


\begin{figure*}[t]
\centering
\includegraphics[width=0.47\textwidth]{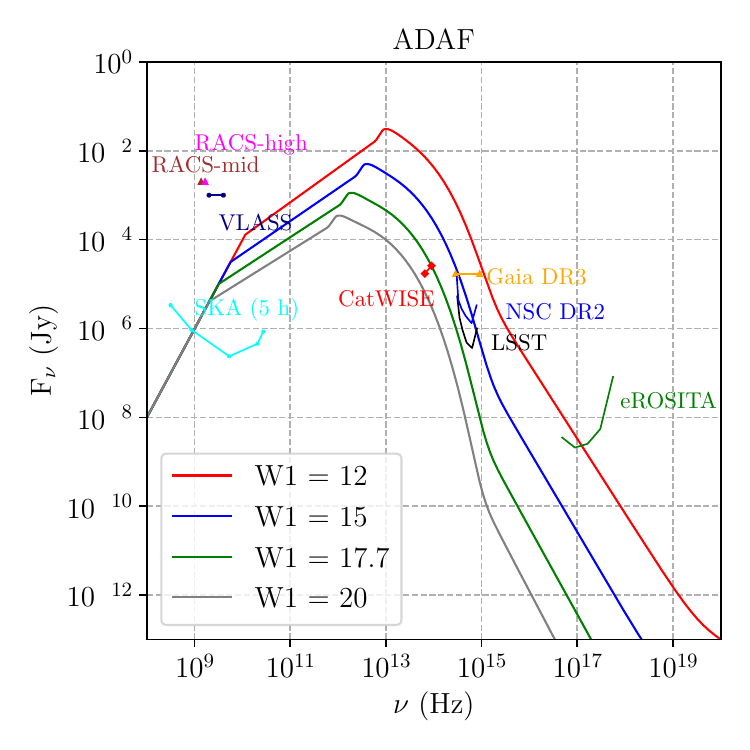}
\includegraphics[width=0.47\textwidth]{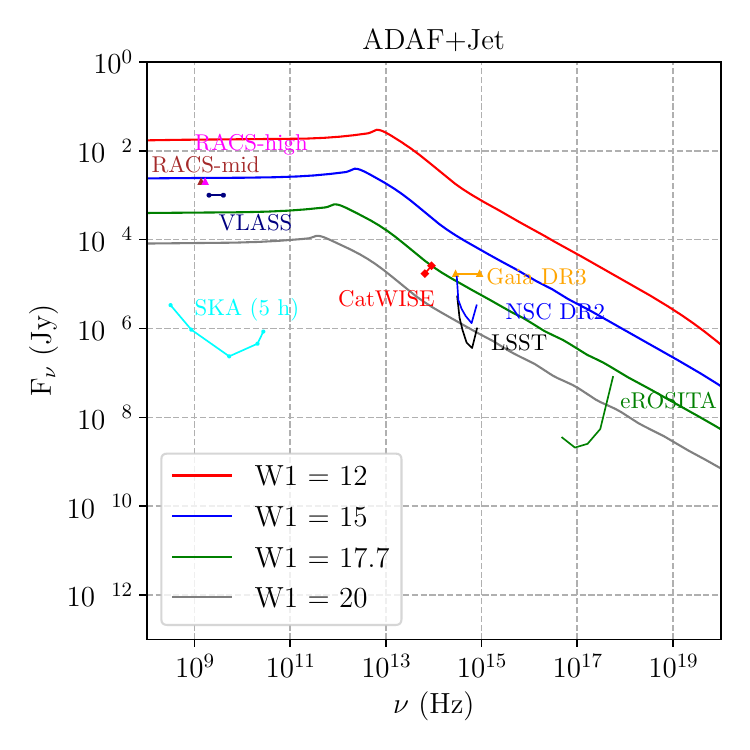}
\vspace{-0.2cm}
\caption{Synthetic spectra of isolated black holes accreting from the interstellar medium in an ADAF model (left panel) and ADAF+Jet model (right panel). The distance of the sources is set to 10~pc and we vary the mass accretion rate such that the red spectra have $W1 = 12$, the blue spectra have $W1 = 15$, the green spectra have $W1 = 17.7$, and the gray spectra have $W1 = 20$. $W1 = 17.7$ is the nominal detection threshold of CatWISE2020, but the catalog includes even fainter sources up to $W1 \approx 20$. In the ADAF+Jet model, we assume that the mass-loss rate into the jet is 10\% and the angle between the jet axis and our line of sight is $20^\circ$ \citep[see][and references therein, for more details]{2026ApJ..1004...21N}. We also report the detection thresholds of RACS-mid (brown), RACS-high (magenta), VLASS~QL1 (dark blue), CatWISE2020 (red), Gaia~DR3 (orange), NSC~DR2 (blue), LSST (black), eROSITA (green), and for a 5~hour observation with SKA (cyan). See the text for more details.}
\label{f-spectra}
\end{figure*}

\section{Spectra of Nearby Black Holes}\label{s-spectra}

Following \citet{2025ApJ...988L..12M}, \citet{2026ApJ..1004...21N} employed the {\tt LLAGNSED} code~\citep{2021ApJ...923..260P} to calculate the spectra of nearby black holes accreting from the interstellar medium. {\tt LLAGNSED} is intended to calculate the spectra of low-luminosity active galactic nuclei based on self-similar ADAF solutions. In the case of a stellar-mass black hole in a warm and partially ionized cloud, the compact object moves through the interstellar medium with a speed exceeding the sound speed of the interstellar medium, and this causes the formation of a Mach cone and a non-isotropic emission of the source. However, there are no models for spectra of such black holes in the literature. The spectra predicted by the {\tt LLAGNSED} model should thus be seen as crude approximations for our case.

Some accretion models predict jet-dominated spectra for very faint systems with total luminosity $L < 10^{-6}~L_{\rm Edd}$, where $L_{\rm Edd}$ is the Eddington luminosity of the source~\citep{2005ApJ...629..408Y}. Such a conclusion seems also to be supported by observations~\citep{2014MNRAS.442L.110X,2016RAA....16...62Y}. \citet{2026ApJ..1004...21N} combined the ADAF spectrum from {\tt LLAGNSED} with the jet model of \citet{2014MNRAS.442L.110X} to calculate possible jet-dominated spectra of black holes accreting from the interstellar medium.

Fig.~\ref{f-spectra} shows the predictions for the spectra of isolated black holes accreting from the interstellar medium in the two scenarios, ADAF (left panel) and ADAF+Jet (right panel), according to the models used in \citet{2026ApJ..1004...21N}. The distance of the sources is set to 10~pc and the mass accretion rate is changed in order to have spectra with $W1 = 12$, 15, 17.7, and 20, where $W1$ is one of the two bands in CatWISE2020 (nominal wavelength 3.4~$\mu$m), $W1 = 17.7$ is the nominal detection threshold of CatWISE2020 in this band, and $W1 \approx 20$ is the maximum sensitivity reached for certain sources. The spectra with $W1 = 12$ and 15 are for reference only. For the ADAF scenario, the corresponding mass accretion rates in Eddington units are $4.5 \cdot 10^{-8}$ ($W1 = 12$), $1.3 \cdot 10^{-8}$ ($W1 = 15$), $4.7 \cdot 10^{-9}$ ($W1 = 17.7$), and $2.1 \cdot 10^{-9}$ ($W1 = 20$). For the ADAF+Jet scenario, we have $2.8 \cdot 10^{-8}$ ($W1 = 12$), $6.8 \cdot 10^{-9}$ ($W1 = 15$), $1.9 \cdot 10^{-9}$ ($W1 = 17.7$), and $5.9 \cdot 10^{-10}$ ($W1 = 20$). For a $10~M_\odot$ black hole in one of the warm and partially ionized clouds nearby the Solar System, we can expect an Eddington-scaled mass accretion rate ranging from $\sim 10^{-8}$ to $\sim 10^{-10}$ \citep[for more details, see][]{2026ApJ..1004...21N}.

Fig.~\ref{f-spectra} also shows the detection thresholds of RACS-mid and RACS-high~\citep{2024PASA...41....3D,2025PASA...42...38D}, VLASS~QL1~\citep{2020PASP..132c5001L}, Gaia~DR3~\citep{2023A&A...674A...1G}, NSC~DR2~\citep{2021AJ....161..192N}, LSST~\citep{2019ApJ...873..111I}, eROSITA~DR1~\citep{2024A&A...682A..34M}, and a 5~hour observation with SKA~\citep{2019arXiv191212699B}. For LSST, we report the single-visit detection threshold because our sources are nearby the Solar System and therefore they are characterized by high proper motion; the first data release is expected in 2028, but prompt products are available since June 30, 2026. For SKA, we report the detection threshold for a 5~hour observation because it is not yet clear the nominal detection threshold of the catalog, which is currently scheduled to be released in 2027.

According to our ADAF model, CatWISE2020 is the catalog that can include the faintest sources and therefore it will be our starting point in the search for nearby black holes. LSST follows immediately after CatWISE2020 (but is not yet available), while NSC~DR2 and Gaia~DR3 can miss some faint isolated black holes present in the CatWISE2020 catalog. On the other hand, eROSITA~DR1 can detect only significantly brighter sources, and RACS-mid, RACS-high, and VLASS~QL1 do not seem to have any detection chance. For SKA, it is difficult to make predictions before the data release, but it promises to be very useful to search for nearby black holes accreting from the interstellar medium.

The conclusions change in the ADAF+Jet model. We note that in Fig.~\ref{f-spectra} we assume that the angle between the jet axis and our line of sight is 20$^\circ$: for a larger angle, the contribution of the jet decreases. For such a favorable condition, our ADAF+Jet model predicts that eROSITA~DR1 is the existing catalog that can catch the faintest sources, followed by CatWISE2020, then NSC~DR2 and Gaia~DR3, and finally the radio catalogs RACS and VLASS~QL1. The upcoming catalogs of LSST and SKA are confirmed as very promising catalogs for the search for these objects.


\section{Search for Nearby Black Holes}\label{s-search}

In the present study, we only consider black holes with an ADAF spectrum and we postpone the ADAF+Jet scenario to the future. The two cases clearly require different strategies. Our ADAF+Jet model predicts that eROSITA is the current observational facility with the best sensitivity; the problem is that its astrometric precision is at the level of a few arcseconds (and even worse for very faint sources), so we cannot easily distinguish the emission of nearby sources.

Our search relies on two important assumptions: $i)$ our theoretical model can approximate well the spectra of black holes accreting from the interstellar medium, and $ii)$ the spectrum of a source does not change appreciably over 15~years, which is roughly the time interval spanned by the observations in the catalogs used in our study. Concerning assumption $i)$, {\tt LLAGNSED} assumes a spherically symmetric ADAF onto a non-rotating black hole, while we expect that our black holes are rotating and move through the interstellar medium with a velocity higher than the sound speed of the interstellar medium. As for assumption $ii)$, we can expect variability in the flux on timescales of order $\tau \sim$~a few~$r_{\rm g}/c \sim 10$~ms for a 10~$M_\odot$ black hole~\citep{2025ApJ...988L..12M}, where $r_{\rm g} = M$ is the gravitational radius of the black hole; such a variability cannot be observed in the catalogs used in our study, as all their flux measurements are averaged over observations of dozens of seconds. Flux variability on timescales of years is possible if the black hole moves to a region with a different particle density.

\begin{figure}[t]
\centering
\includegraphics[width=0.45\textwidth,trim={8.5cm 0cm 1cm 0cm},clip]{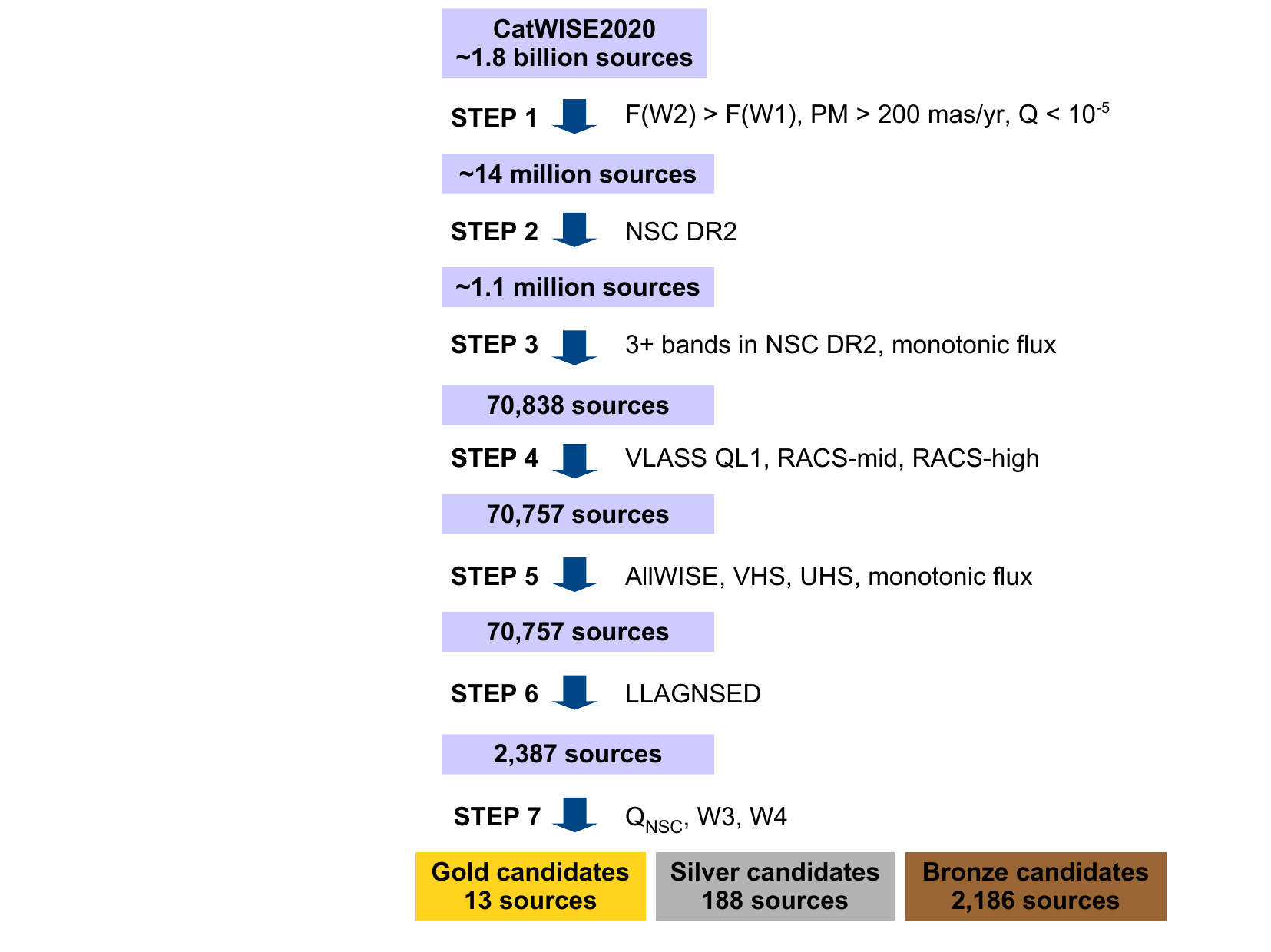}
\caption{Flowchart for the selection of black hole candidates with spectra compatible with the ADAF model.}
\label{f-fc}
\end{figure}

Fig.~\ref{f-fc} shows the flowchart for the selection of our candidates. We start from CatWISE2020~\citep{2020ApJS..247...69E,2021ApJS..253....8M}, because according to our ADAF model it is the most suitable available catalog to detect faint black holes accreting from the interstellar medium. CatWISE2020 has about 1.8~billion sources. We require that our sources have $F(W2) > F(W1)$, where $F(Wi)$ is the flux measurement of the source in the $Wi$ band of CatWISE2020; indeed, our model predicts a monotonically decreasing flux in the IR and at higher frequencies. We also require that the sources have proper motion ${\rm PM} > 200$~mas/year: for a source at 10~pc (15~pc), this corresponds to a transverse velocity of $\sim$9.5~km/s ($\sim$14~km/s). While we do not know the velocity distribution of isolated black holes, we can expect the typical black hole velocity is between the velocity dispersion of nearby massive stars $v_{\rm mstar} \sim 20$~km/s and $v_{\rm BH,kick} \sim 60$~km/s~\citep{2013MNRAS.430.1538F}, where $v_{\rm BH,kick}$ is inferred by rescaling the velocity distribution measured for neutron stars to a 10~$M_\odot$ object (i.e., we assume that black holes and neutron stars receive comparable kicks during the supernova explosion).

Since the proper motion measurement of many faint sources in CatWISE2020 is affected by large uncertainties, we need to select the sources with decent measurements. We impose $Q < 10^{-5}$, where $Q$ is 
\be\label{eq-Q}
Q &=& e^{- \chi^2_{\rm PM}/2} \, , \nonumber\\
\chi^2_{\rm PM} &=& \left(\frac{{\rm PM}_{\rm RA}}{\sigma_{\rm RA}}\right)^2 
+ \left(\frac{{\rm PM}_{\rm Dec}}{\sigma_{\rm Dec}}\right)^2 \, ,
\ee
${\rm PM}_{\rm RA}$ is the right ascension proper motion of the source, ${\rm PM}_{\rm Dec}$ is the declination proper motion of the source, and $\sigma_{\rm RA}$ and $\sigma_{\rm Dec}$ are their uncertainties. This is equivalent to require that the measured proper motion vector has a Mahalanobis distance from zero proper motion of $D_{\rm M} > 4.80$, where $D_{\rm M}^2 = \chi^2_{\rm PM}$. Under these conditions (step~1 in Fig.~\ref{f-fc}), we get a list of about 14~million sources.

At step~2, we cross-match our source list with the optical catalog NSC~DR2~\citep{2021AJ....161..192N}. 
Our cross-matching method with NSC~DR2 and with other catalogs used later is as follows. The CatWISE2020 positions are propagated to the reference epochs of the other catalog. The corresponding positional covariance matrix is propagated using the Jacobian transformation ($C(t)=JC_0J^T$), incorporating uncertainties in both position and proper motion~\citep{2014A&A...570A..62B}. The positional consistency of each candidate counterpart is then evaluated using the squared Mahalanobis distance, calculated from the positional offsets and the sum of the propagated CatWISE2020 covariance and the external catalog covariance~\citep{2019A&A...621A.144M}. We adopt a threshold of $\chi^2_{\mathrm{pos}} \leq 28.74$ for optical and IR catalogs, corresponding to a two-dimensional 5-$\sigma$ equivalent confidence region, and $\chi^2_{\mathrm{pos}} \leq 9.21$ for radio catalogs, corresponding to a 99\% confidence level. When we cross-match the 14~million sources with NSC~DR2, we also require that proper motions in CatWISE2020 and NSC~DR2 are consistent at the 99\% confidence level. At the end of step~2, we have about 1.1~million sources.

Unfortunately, many faint sources have only a flux measurement in one or two bands of NSC~DR2. While we cannot exclude the possibility that some of these sources with only one or two flux measurements are black holes, we decide to consider only those sources with a flux measurement in at least three bands of NSC~DR2. We also impose that the flux of every source monotonically decreases with frequency, as this seems to be quite a robust prediction of our theoretical model\footnote{The spectra of black holes accreting from the interstellar medium in the IR and optical bands should be dominated by synchrotron emission: while the slope of the spectrum may depend on the exact model, we can unlikely have bumps.}. With these two requirements (step~3), we remain with 70,838~sources.

At step~4, we cross-match our sources with the radio catalogs VLASS~QL1~\citep{2020PASP..132c5001L}, RACS-mid, and RACS-high~\citep{2024PASA...41....3D,2025PASA...42...38D}. Since our theoretical model predicts a radio flux significantly lower than the detection thresholds of these catalogs, we remove all sources with radio emission. Our list of black hole candidates now counts 70,575~sources.

To further increase the number of bands with a flux measurement of our sources, we cross-match our list with the IR catalogs AllWISE~\citep{2013wise.rept....1C}, VHS~\citep{2013Msngr.154...35M} and UHS~\citep{2018MNRAS.473.5113D,2025AJ....170...86S}, but we do not remove the sources without cross-matching in these catalogs (step~5). AllWISE has four bands: W1, W2, W3, and W4. When we cross-match the sources, we require that the W1 and W2 magnitude measurements in CatWISE2020 and AllWISE are consistent within 3-$\sigma$. If a source appears in AllWISE, we use this catalog only for the flux measurements in the W3 and W4 bands, while we use the measurements reported in CatWISE2020 for the W1 and W2 bands because more precise. We note that the detection thresholds of VHS and UHS are a bit higher than the flux values expected in our model, so the vast majority of our candidates do not have any counterpart in VHS or UHS. At the end of the cross-matching, we have the same list of sources as after step~4, but now some sources have even flux measurements (or upper limits) in the W3 and W4 bands in AllWISE, VHS, and UHS. We impose again that the flux of every source monotonically decreases with frequency, but there are no sources with clear bumps in the spectra, so we remain with 70,575~sources.

We fit the spectra of these 70,575~sources with {\tt LLAGNSED} with two free parameters: the mass accretion rate and the distance (step~6). For every source, we calculate the minimum of $\chi^2$ and its corresponding lower-tail probability, defined as
\be
p_{\rm low} = P \left( \chi^2_\nu \le \chi^2_{\rm min} \right) \, ,
\ee
where $\nu = N - k$ is the number of degrees of freedom, $N$ is the number of observed data points, and $k$ is the number of free parameters ($k = 2$ in our case, the mass accretion rate and the distance). This probability represents the likelihood of obtaining a $\chi^2$ value equal to or smaller than the observed value, assuming that the model correctly describes the data and that the observational uncertainties are accurately estimated. We additionally express this probability in terms of a one-sided Gaussian-equivalent significance
\be
Z = \Phi^{-1} \left( 1 - p_{\rm low} \right) \, ,
\ee
where $\Phi$ denotes the cumulative distribution function of the standard normal distribution. Considering the large uncertainties in our theoretical model and that the flux uncertainties of our candidates are likely underestimated because they are all very faint sources around the nominal detection thresholds of our catalogs, we decided to keep all sources with a $p_{\rm low}$ below $1 - 10^{-16}$. We remain with 2,387~sources.

We realize that most of these 2,387~sources are likely $i)$ distant faint sources with spurious proper motion (despite our requirement $Q < 10^{-5}$ at step~1), and $ii)$ cool brown dwarfs. If they are blended reddened stars, the spurious motion measurement can be caused by the blending. There may even be faint background galaxies, where the faint, noise-dominated measurements near catalog detection limits have resulted in spurious proper motions in CatWISE2020. In order to remove distant sources with spurious proper motion, we calculate $Q$ in Eq.~(\ref{eq-Q}) from the proper motion measurements in NSC~DR2 (we denote this quantity $Q_{\rm NSC}$), and we require $Q_{\rm NSC} < 0.1$. To remove brown dwarfs from our list, we require a flux measurement in the W3 and/or W4 bands. We thus group our sources into three classes (step~7):

{\bf ``Gold'' candidates}: this is the list of our most promising black hole candidates. They have $Q_{\rm NSC} < 0.1$ and a flux measurement in the W3 or W4 bands. The condition $Q_{\rm NSC} < 0.1$ can help remove distant sources with spurious proper motion and the condition of a flux measurement in the W3 or W4 bands can help remove brown dwarfs. We have 13~sources in this class; 9~sources have a flux measurement in the W3 band and not in the W4 band, and 4~sources have a flux measurement in the W4 band and not in the W3 band.

{\bf ``Silver'' candidates}: this sources have $Q_{\rm NSC} < 0.1$ but no flux measurement in the W3 or W4 bands; some sources have upper limits in the W3 and W4 bands and some sources are not in AllWISE at all. Most of the sources are likely brown dwarfs, but we cannot exclude that they are black holes either, because the detection thresholds in the W3 ad W4 bands of AllWISE are above the fluxes expected for a black hole accreting from the interstellar medium. We have 188~sources in this class.

{\bf ``Bronze'' candidates}: this is the list of weak black hole candidates because we cannot say whether they are nearby sources or distant sources with spurious proper motion, since $Q_{\rm NSC} > 0.1$. If they are nearby sources, they may be brown dwarfs, because we just have a flux monotonically decreasing from W2 to the optical bands of NSC~DR2, normally without any flux measurement in VHS or UHS. We have 2,186~sources in this class.


\begin{table*}[t]
\centering
\renewcommand\arraystretch{1.2}
\caption{List of the 13~gold black hole candidates. For every candidate, we report: source ID in CatWISE2020, source position (RA and Dec) in CatWISE2020, proper motion measurement in CatWISE2020, source ID in NSC~DR2, proper motion measurement in NSC~DR2, and significance $Z$. Positions are in degrees and proper motions are in mas/yr. The reported proper motion uncertainties correspond to 1-$\sigma$ limit.}
\label{t-gold}
{\scriptsize
\begin{tabular}{c|lll|ll|c}
\hline
\hline
 & \multicolumn{3}{c}{CatWISE2020} & \multicolumn{2}{c}{NSC~DR2} & $Z$ \\
 & Source ID & Position (deg) & PM (mas/yr) & Source ID & PM (mas/yr) & \\
\hline
1 & J121921.12$-$311501.3 & ${\rm RA} = 184.8380384$ & ${\rm PM}_{\rm RA} = 91.64 \pm 31.10$ & 148999\_2739 & ${\rm PM}_{\rm RA} = 65.03 \pm 42.37$ & $-1.175$ \\
 & & ${\rm Dec} = -31.2503368$ & ${\rm PM}_{\rm Dec} = 210.47 \pm 33.90$ & & ${\rm PM}_{\rm Dec} = 109.70 \pm 41.93$ & \\
\hline
2 & J121906.33$-$303453.7 & ${\rm RA} = 184.7764281$ & ${\rm PM}_{\rm RA} = 156.82 \pm 85.20$ & 148486\_14300 & ${\rm PM}_{\rm RA} = -90809.13 \pm 50604.03$ & $-2.361$ \\
 & & ${\rm Dec} = -30.5815028$ & ${\rm PM}_{\rm Dec} = 402.72 \pm 87.60$ & & ${\rm PM}_{\rm Dec} = 66558.94 \pm 50460.73$ & \\
\hline
3 & J021440.70+065012.4 & ${\rm RA} = 33.6695748$ & ${\rm PM}_{\rm RA} = 201.53 \pm 55.90$ & 86320\_1198 & ${\rm PM}_{\rm RA} = 46.76 \pm 142.68$ & $-5.192$ \\
 & & ${\rm Dec} = 6.8367332$ & ${\rm PM}_{\rm Dec} = 240.79 \pm 58.20$ & & ${\rm PM}_{\rm Dec} = 473.27 \pm 149.96$ & \\
\hline
4 & J225437.13$-$114503.9 & ${\rm RA} = 343.6547104$ & ${\rm PM}_{\rm RA} = 266.25 \pm 50.10$ & 118505\_4297 & ${\rm PM}_{\rm RA} = 153.89 \pm 159.47$ & $-5.729$ \\
 & & ${\rm Dec} = -11.7511045$ & ${\rm PM}_{\rm Dec} = -62.78 \pm 46.80$ & & ${\rm PM}_{\rm Dec} = -359.73 \pm 160.36$ & \\
\hline
5 & J222404.65$-$133446.4 & ${\rm RA} = 336.0193729$ & ${\rm PM}_{\rm RA} = 214.84 \pm 51.00$ & 121566\_3243 & ${\rm PM}_{\rm RA} = 374.29 \pm 121.72$ & $-5.743$ \\
 & & ${\rm Dec} = -13.5795478$ & ${\rm PM}_{\rm Dec} = 176.36 \pm 53.20$ & & ${\rm PM}_{\rm Dec} = -93.90 \pm 121.97$ & \\
\hline
6 & J100641.45+245940.4 & ${\rm RA} = 151.6727173$ & ${\rm PM}_{\rm RA} = 326.73 \pm 72.80$ & 56792\_4846 & ${\rm PM}_{\rm RA} = 120.03 \pm 32.66$ & $-6.183$ \\
 & & ${\rm Dec} = 24.9945688$ & ${\rm PM}_{\rm Dec} = 141.87 \pm 77.30$ & & ${\rm PM}_{\rm Dec} = 36.76 \pm 33.21$ & \\
\hline
7 & J120545.68$-$351238.3 & ${\rm RA} = 181.4403218$ & ${\rm PM}_{\rm RA} = -117.93 \pm 55.80$ & 155138\_13364 & ${\rm PM}_{\rm RA} = 52.98 \pm 48.61$ & $-6.363$ \\
 & & ${\rm Dec} = -35.2105798$ & ${\rm PM}_{\rm Dec} = 319.69 \pm 63.00$ & & ${\rm PM}_{\rm Dec} = 258.89 \pm 48.34$ & \\
\hline
8 & J155901.92+053805.7 & ${\rm RA} = 239.7579877$ & ${\rm PM}_{\rm RA} = 274.76 \pm 54.90$ & 88661\_4753 & ${\rm PM}_{\rm RA} = 154.17 \pm 45.08$ & $-6.614$ \\
 & & ${\rm Dec} = 5.6349363$ & ${\rm PM}_{\rm Dec} = -67.20 \pm 66.90$ & & ${\rm PM}_{\rm Dec} = -7.56 \pm 45.33$ & \\
\hline
9 & J034646.95$-$323350.7 & ${\rm RA} = 56.6955675$ & ${\rm PM}_{\rm RA} = -231.14 \pm 60.50$ & 151376\_5958 & ${\rm PM}_{\rm RA} = -54.78 \pm 58.25$ & $-6.804$ \\
 & & ${\rm Dec} = -32.5640696$ & ${\rm PM}_{\rm Dec} = 172.70 \pm 58.10$ & & ${\rm PM}_{\rm Dec} = 125.78 \pm 58.26$ & \\
\hline
10 & J004410.18+200425.1 & ${\rm RA} = 11.0425614$ & ${\rm PM}_{\rm RA} = 289.54 \pm 57.00$ & 64271\_5205 & ${\rm PM}_{\rm RA} = 251.30 \pm 109.84$ & $-7.009$ \\
 & & ${\rm Dec} = 20.0735832$ & ${\rm PM}_{\rm Dec} = -254.54 \pm 58.20$ & & ${\rm PM}_{\rm Dec} = -83.75 \pm 110.33$ & \\
\hline
11 & J003251.14+234406.4 & ${\rm RA} = 8.2130988$ & ${\rm PM}_{\rm RA} = -172.06 \pm 54.60$ & 58636\_4029 & ${\rm PM}_{\rm RA} = -1.62 \pm 68.14$ & $-7.409$ \\
 & & ${\rm Dec} = 23.7351297$ & ${\rm PM}_{\rm Dec} = 198.89 \pm 51.20$ & & ${\rm PM}_{\rm Dec} = 254.13 \pm 68.31$ & \\
\hline
12 & J111304.67$-$045649.1 & ${\rm RA} = 168.2694548$ & ${\rm PM}_{\rm RA} = -96.34 \pm 35.50$ & 106479\_6000 & ${\rm PM}_{\rm RA} = 29.02 \pm 44.11$ & $-7.928$ \\
 & & ${\rm Dec} = -4.9470064$ & ${\rm PM}_{\rm Dec} = 212.37 \pm 40.90$ & & ${\rm PM}_{\rm Dec} = 96.47 \pm 44.17$ & \\
\hline
13 & J234032.07$-$032711.1 & ${\rm RA} = 355.1337371$ & ${\rm PM}_{\rm RA} = -616.05 \pm 128.10$ & 104185\_5969 & ${\rm PM}_{\rm RA} = -19814.77 \pm 8706.13$ & $-7.992$ \\
 & & ${\rm Dec} = -3.4530369$ & ${\rm PM}_{\rm Dec} = -260.94 \pm 115.30$ & & ${\rm PM}_{\rm Dec} = -7457.28 \pm 8732.69$ & \\
\hline
\hline
\end{tabular}
}
\vspace{0.8cm}
\end{table*}

\section{Discussion and Conclusions}\label{s-d-c}

The search for nearby black holes presented in the previous section has led to the identification of 13~gold candidates, 188~silver candidates, and 2,186~bronze candidates. The goal of our study is not to provide the full list of potential candidates in existing catalogs but to outline a strategy for the search for nearby black holes accreting from the interstellar medium. The challenge is that these objects are expected to be very faint sources. We need to impose some quality requirements (e.g., requirements on $Q$, the minimum number of detection bands, etc.) because, otherwise, we would obtain a very long list of potential candidates with poor measurements. On the other hand, we cannot impose too stringent requirements on the quality of these measurements, because nearby black holes are certainly very faint sources and we risk missing them. These selection criteria can certainly be improved in future studies.

\begin{figure*}[t]
\centering
\includegraphics[width=0.99\textwidth]{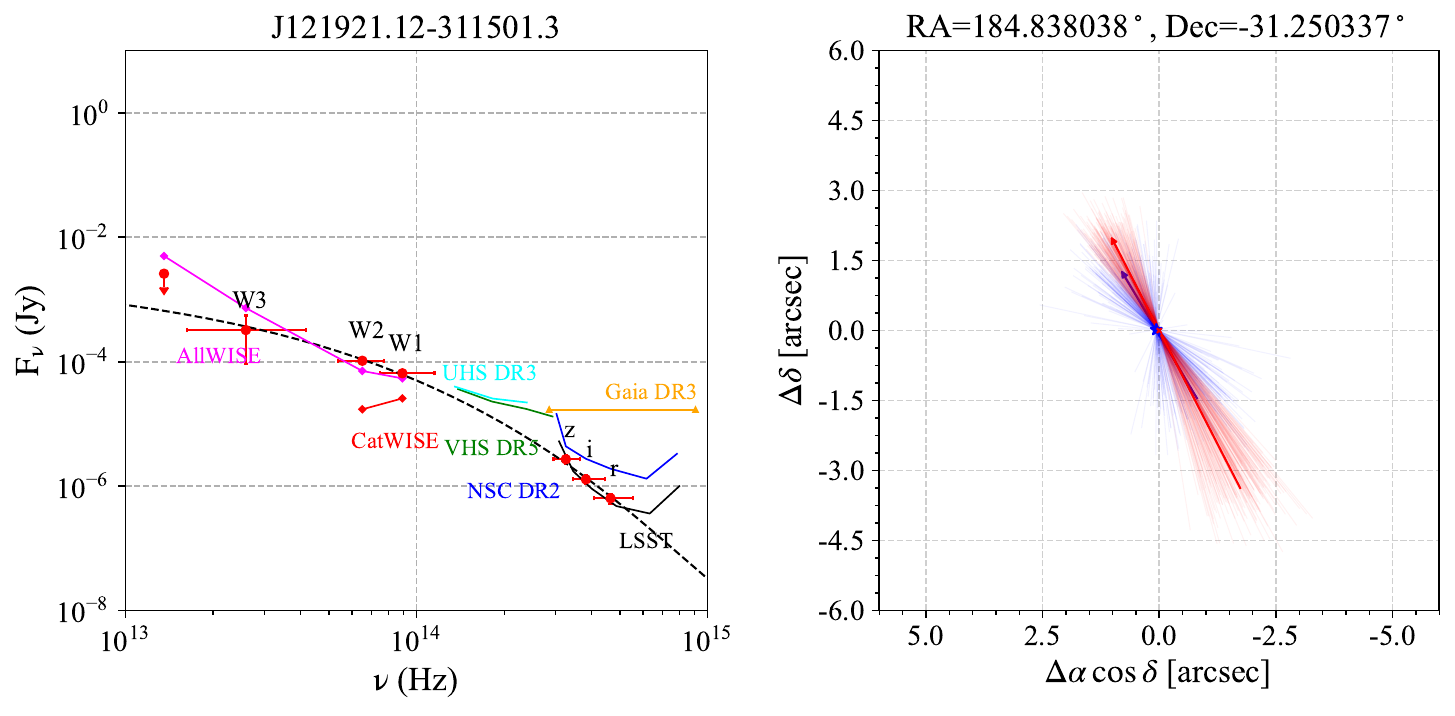}
\caption{Gold candidate J121921.12$-$311501.3. Left panel: spectrum of the source. The red dots are the flux measurements; their vertical error bars represent the 2-$\sigma$ flux uncertainties, while their horizontal error bars indicate the band widths. The 2-$\sigma$ flux uncertainties in the r, i, z, W1, and W2 bands are not visible because the error bars are so small that they are within the red dots. The black-dashed curve is the best-fit of {\tt LLAGNSED}. The plot also shows the nominal detection thresholds of AllWISE (magenta), CatWISE2020 (red), UHS~DR3 (cyan), VHS~DR5 (green), Gaia~DR3 (orange), NSC~DR2 (blue), and LSST (black). Right panel: map of the sky region around the source. The location of the source in CatWISE2020 is indicated by the red cross at the center of the plot $\Delta\delta = \Delta\alpha\cos\delta = 0$ and the red arrow shows its predicted trajectory in the sky. The location of the source in NSC~DR2 is indicated by the blue star and the blue arrow shows its predicted trajectory in the sky. The tips of the arrows are at the predicted positions of the source in 2025 and the opposite ends are at the predicted positions of the source in 2000. The thin segments without tips indicate the 1-$\sigma$ uncertainties of these trajectories (red segments for CatWISE2020 and blue segments for NSC~DR2): they are randomly sampled from a two-dimensional Gaussian distribution centered on the measured proper motion. We note that there are no other sources reported in CatWISE2020, NSC~DR2, Gaia~DR3, VHS~DR5, VLASS~QL1, RACS-mid, RACS-high, eROSITA~DR2, and SIMBAD in this region of the sky.}
\label{f-gold}
\end{figure*}

The list of 13~gold black hole candidates is reported in Tab.~\ref{t-gold}. Fig.~\ref{f-gold} shows the spectrum of the gold candidate J121921.12$-$311501.3 and the sky region around this source; assuming a $10~M_\odot$ black hole, the Eddington-scaled mass accretion rate of the best-fit is $6.9 \cdot 10^{-9}$. The spectra and the sky regions of all gold candidates are presented in the Supplemental Material. All our candidates are very faint sources, with fluxes close to, and often below, the nominal detection thresholds of AllWISE and NSC~DR2. Since they are very faint, they do not have reliable parallax measurements in CatWISE2020. We checked whether our gold candidates are in Gaia~DR3, SIMBAD~\citep{2000A&AS..143....9W}, or UltraCoolSheet~\citep{best_2025_15802304}, but we did not find any source with our cross-matching method described in Section~\ref{s-search}\footnote{We have also checked whether our silver or bronze candidates are in Gaia~DR3, SIMBAD, or UltraCoolSheet. With our cross-matching method, we found 3~SIMBAD sources potentially associated with 3~bronze candidates and classified as AGN in SIMBAD. We also found 37~Gaia sources potentially associated with 37~bronze candidates. However, most of these sources are near the Galactic plane, which is a crowded region where it is challenging to associate sources from different catalogs. We do not investigate these sources further because at this stage we consider all bronze sources as very weak candidates.}. This also means that all these sources have no classification in the existing literature.

Even if our gold candidates have $Q_{\rm NSC} < 0.1$ and a flux measurement in the W3 or W4 bands, this is not enough to be sure that we have removed all distant sources with spurious proper motion and brown dwarfs. More robust measurements of the proper motions are necessary to exclude distant sources, because the proper motion measurements in NSC~DR2 can also be incorrect and accidentally consistent with those in CatWISE2020. Moreover, $Q_{\rm NSC} < 0.1$ is quite a weak requirement. Nearby sources with a maximum in the W3 band can still be cool brown dwarfs, so a flux measurement in the W3 band is not sufficient to rule out brown dwarfs. Concerning the 4~gold sources with flux measurements in the W4 band, assuming that they are really nearby sources instead of distant objects, we can exclude they are ``bare'' brown dwarfs. Still they may be nearby low-mass stars or brown dwarfs with cool disks rather than black holes accreting from the interstellar medium.

\subsection{Candidate confirmation}

As a conclusion of our work, we can wonder how we can confirm or reject our gold candidates. Unfortunately, the SKA catalog is not yet available (it is currently expected to be released in 2027): as shown in Fig.~\ref{f-spectra}, our theoretical models predict that SKA can be very helpful to identify nearby black holes accreting from the interstellar medium. Moreover, the SKA catalog is supposed to report the polarization measurements of its sources: for a black hole accreting from the interstellar medium, we expect that the radio emission is partially linearly polarized because it is produced by synchrotron processes. Such a prediction can help distinguishing black holes from other sources.

Without the catalogs available in the near future, the confirmation of our gold candidates can proceed along the following steps:

{\bf Step~1: astrometric confirmation}. For the moment, none of our gold candidates has a reliable parallax measurement. We could measure the parallax of our sources with LSST, but this would be challenging with the data collected over only $\sim$1~year: the LSST single-visit astrometric precision is $\sim$10~mas for very faint sources, and we have to measure both the parallax and the proper motion of a source. 8-10~m ground-based telescopes -- e.g., Keck, Very Large Telescope (VLT), or Gemini -- could do the job, but this would require submitting a proposal to quite competitive observational facilities.

{\bf Step~2: radio emission confirmation}. Without the SKA catalog, the simplest way to confirm the radio emission of our candidates is to observe our sources with a radio interferometer such as the Karl G. Jansky Very Large Array (VLA). We need to be able to detect sources with a flux of $\sim$10~$\mu$Jy at $\sim$10~GHz and with arcsecond or sub-arcsecond astrometric precision, which is achievable by VLA. Concerning Chinese observational facilities, we could use the Five-hundred-meter Aperture Spherical radio Telescope (FAST) in conjunction with the Chinese VLBI Network (e.g., Shanghai Tianma 65~m, Xinjiang Nanshan 26~m, and Shaanxi Haoping 40~m), which would permit to detect sources with a flux of $\sim$1~$\mu$Jy at $\sim$1~GHz.

{\bf Step~3: spectroscopy confirmation}. The spectrum of a black hole accreting from the interstellar medium is expected to be quite smooth, without particular features. On the contrary, brown dwarfs and dusty stars are characterized by strong absorption lines in the optical and near-IR bands. Distant galaxies have spectra with absorption and emission lines. 8-10~m ground-based telescopes such as Keck, VLT, or Gemini can efficiently get spectra for faint targets such as our candidates, and the chances that an observational proposal is accepted are higher than with the James Webb Space Telescope. However, the competition level for these observational facilities is higher than that for VLA, so this suggests getting the spectroscopy confirmation after the radio emission one.

{\bf Step~4: submillimeter emission confirmation}. Our theoretical model predicts that the spectrum of a black hole accreting from the interstellar medium peaks in the submillimeter range. This could be tested by an observational facility like the Atacama Large Millimeter/submillimeter Array (ALMA). Since the observing time of ALMA is more difficult to obtain than Keck, VLT, or Gemini, we leave this check as the last step. For a black hole accreting from the interstellar medium, we can expect a flux of $\sim$1~mJy at 1~THz, which can be easily detected by ALMA Band~9 (650~GHz) and Band~10 (up to 950~GHz) with a short observation.


\vspace{0.5cm}

{\bf Acknowledgments --}
We thank Artur Avakyan for reading a preliminary version of this manuscript and providing useful feedback.
The work of A.N., C.B., Y.F., and L.G. was supported by the National Natural Science Foundation of China (NSFC), Grant No.~W2531002, and by the National Key R\&D Program of the Ministry of Science and Technology of China, Grant No.~2026YFE0110900.
A.N. also acknowledges the support from the Shanghai Government Scholarship (SGS).
F.G.X. is supported in part by the National Natural Science Foundation of China (NSFC), Grant No.~12373017, and the State Key Laboratory of Radio Astronomy and Technology (Chinese Academy of Sciences).

\vspace{0.3cm}

{\bf Author Contribution --} 
A.N. selected and analyzed the sources, and created all plots in the article and in the Supplemental Material. C.B. conceived and supervised the project, and wrote the first version of the manuscript. All authors provided feedback on interpretations of the results and the manuscript.

\bibliographystyle{aa}
\bibliography{references}

\clearpage
\newpage
\onecolumngrid
\begin{center}
\textbf{Supplemental Material:\\
A Multi-Wavelength Search for Nearby Black Holes Accreting from the Interstellar Medium}
\medskip\\
{Abdurakhmon~Nosirov, Cosimo~Bambi, Andrea~Santangelo, Yi~Fan, Leda~Gao, Jos~de~Bruijne, Ningyue~Fan, Jiachen~Jiang, Federico~Marocco, and Fu-Guo~Xie}
\end{center}

\renewcommand{\thesection}{S\arabic{section}}
\renewcommand{\theequation}{S\arabic{equation}}
\renewcommand{\thefigure}{S\arabic{figure}}
\renewcommand{\thetable}{S\arabic{table}}
\renewcommand{\thepage}{S\arabic{page}}
\setcounter{equation}{0}
\setcounter{figure}{0}
\setcounter{table}{0}
\setcounter{page}{1}
\setcounter{secnumdepth}{4}

\section*{Gold Candidates}

\begin{figure*}[t]
\centering
\includegraphics[width=0.95\textwidth]{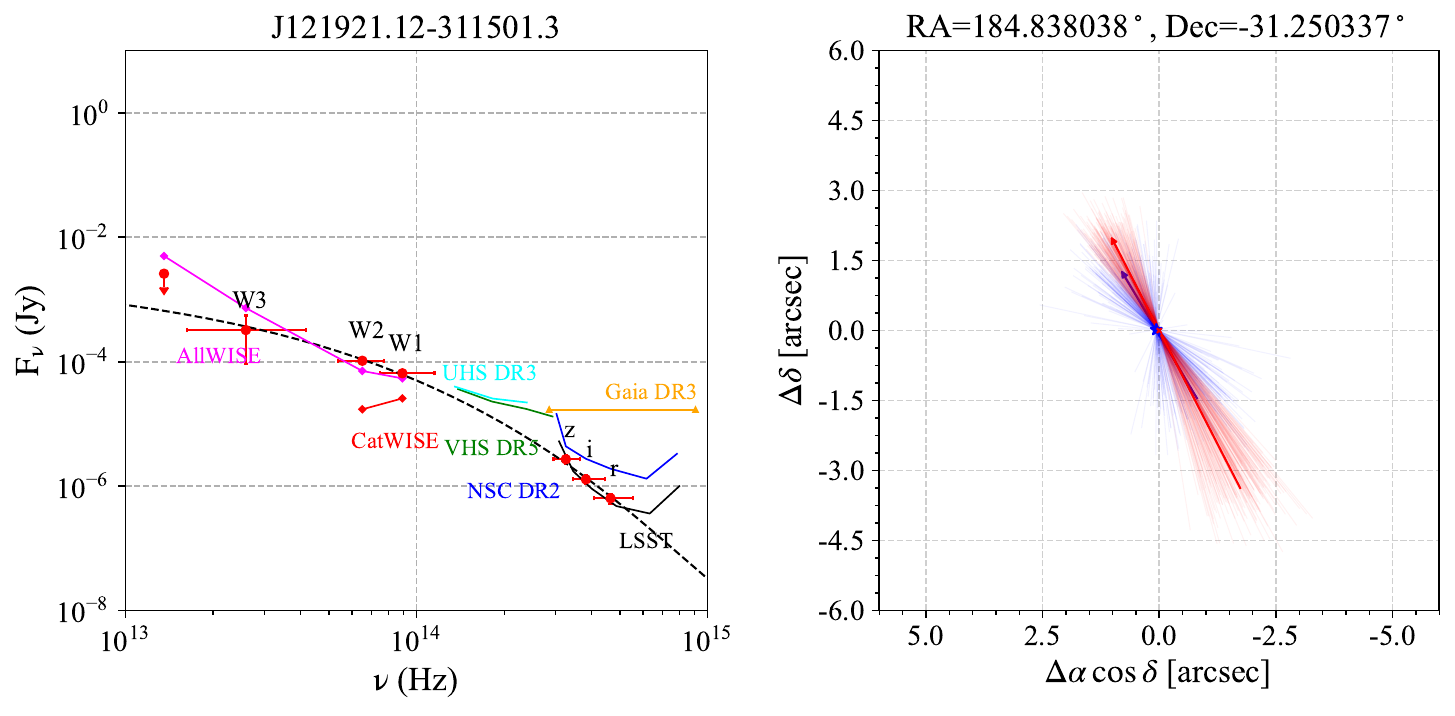}
\caption{Source J121921.12$-$311501.3.}
\label{f-s1}
\end{figure*}

\begin{figure*}[t]
\centering
\includegraphics[width=0.95\textwidth]{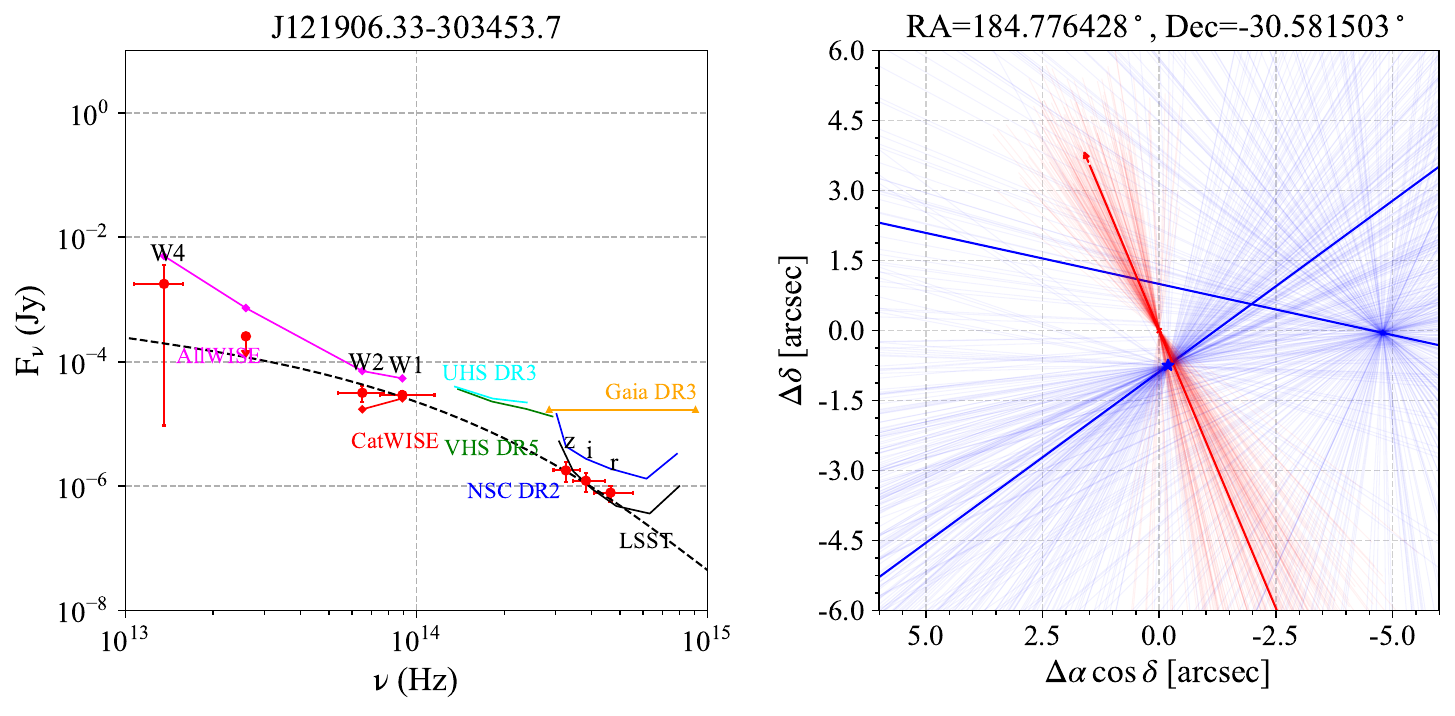}
\caption{Source J121906.33$-$303453.7.}
\label{f-s2}
\end{figure*}

\begin{figure*}[t]
\centering
\includegraphics[width=0.95\textwidth]{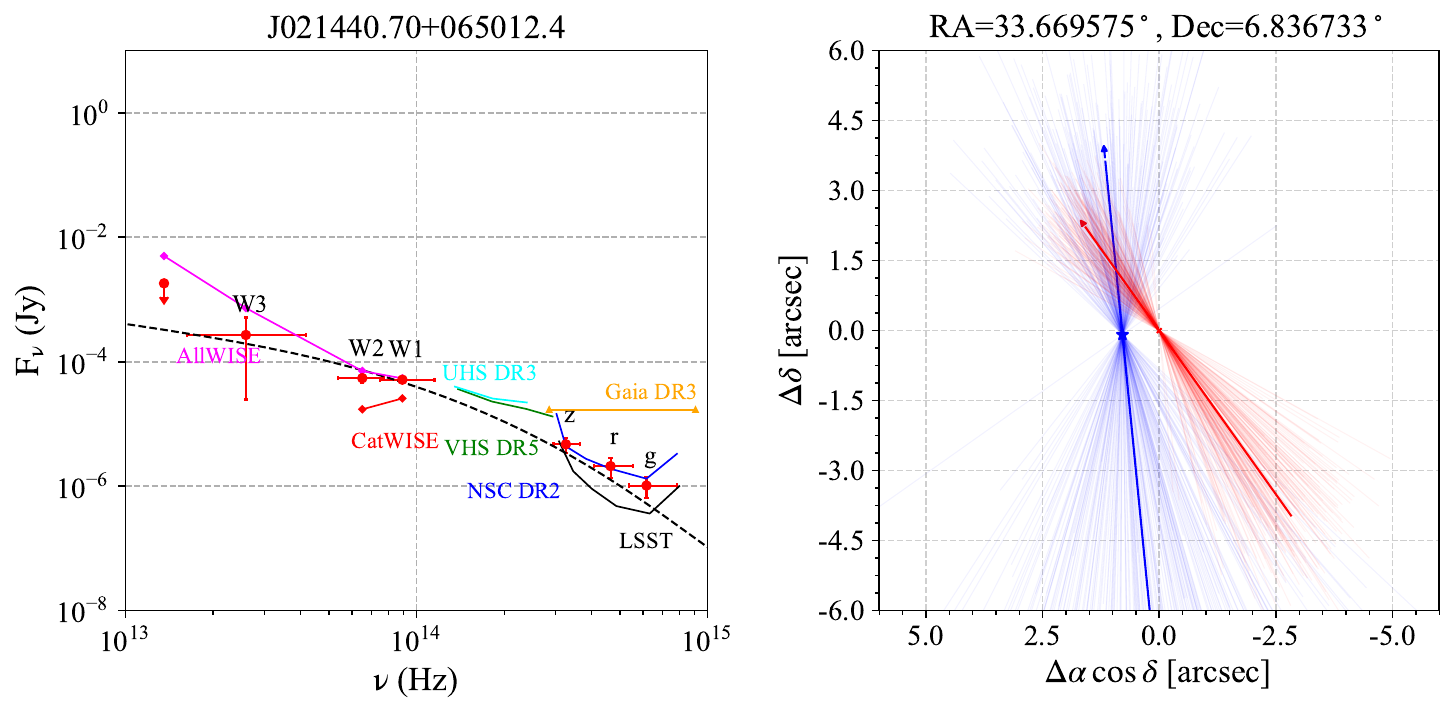}
\caption{Source J021440.70+065012.4.}
\label{f-s3}
\end{figure*}

\begin{figure*}[t]
\centering
\includegraphics[width=0.95\textwidth]{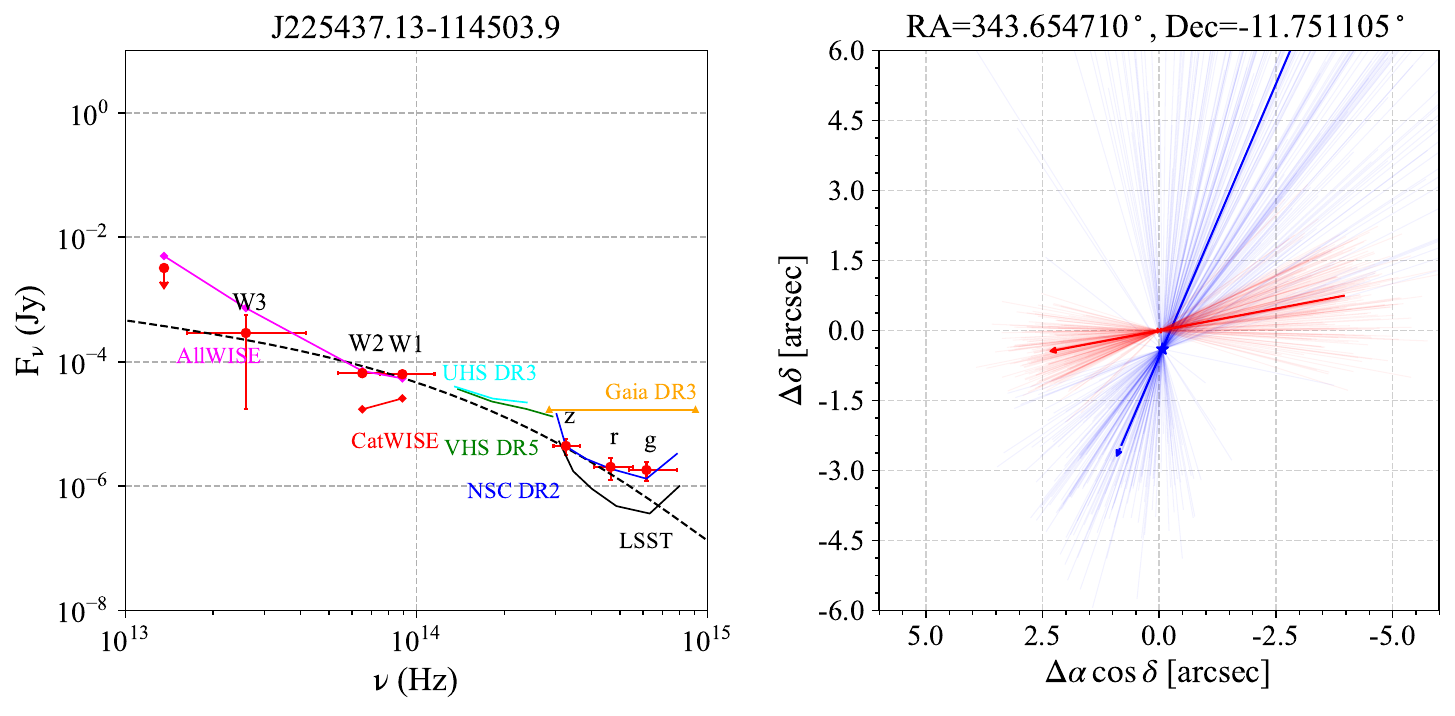}
\caption{Source J225437.13$-$114503.9.}
\label{f-s4}
\end{figure*}

\begin{figure*}[t]
\centering
\includegraphics[width=0.95\textwidth]{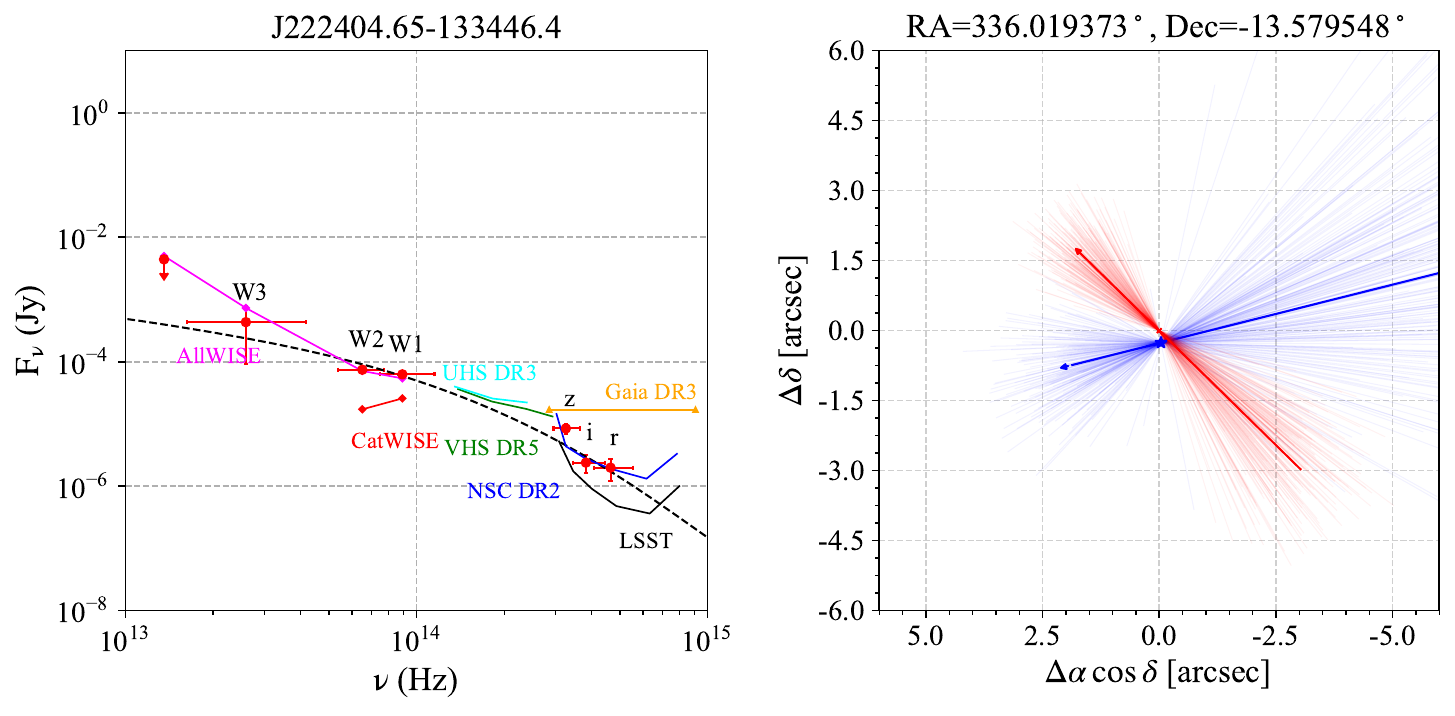}
\caption{Source J222404.65$-$133446.4.}
\label{f-s5}
\end{figure*}

\begin{figure*}[t]
\centering
\includegraphics[width=0.95\textwidth]{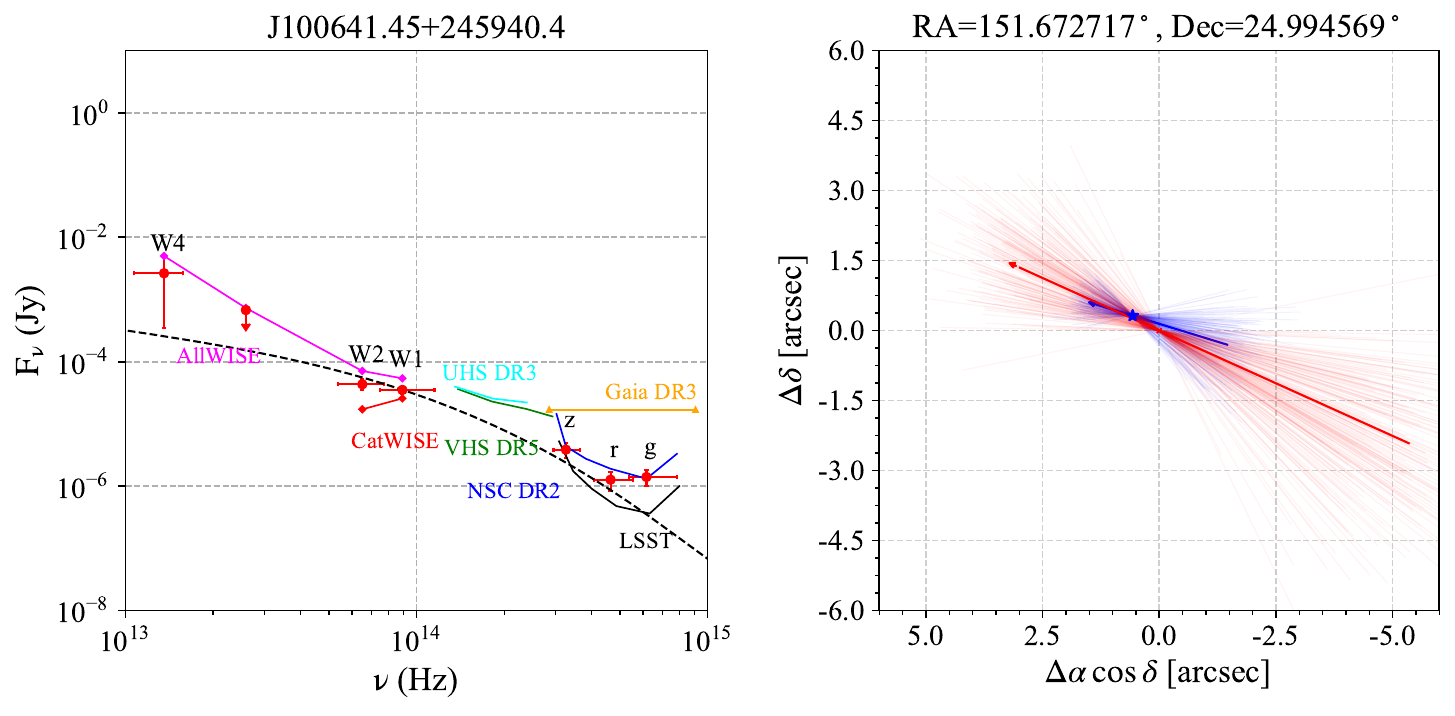}
\caption{Source J100641.45+245940.4.}
\label{f-s6}
\end{figure*}

\begin{figure*}[t]
\centering
\includegraphics[width=0.95\textwidth]{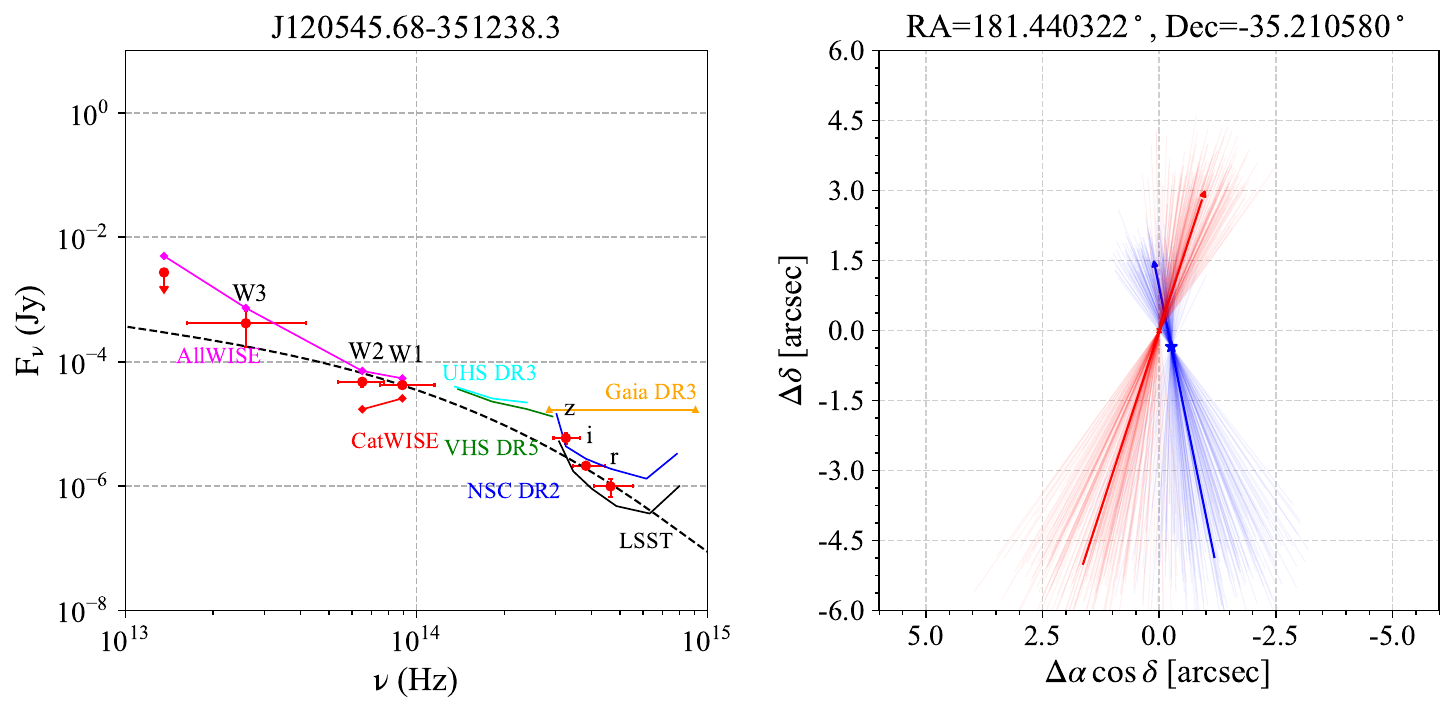}
\caption{Source J120545.68$-$351238.3.}
\label{f-s7}
\end{figure*}

\begin{figure*}[t]
\centering
\includegraphics[width=0.95\textwidth]{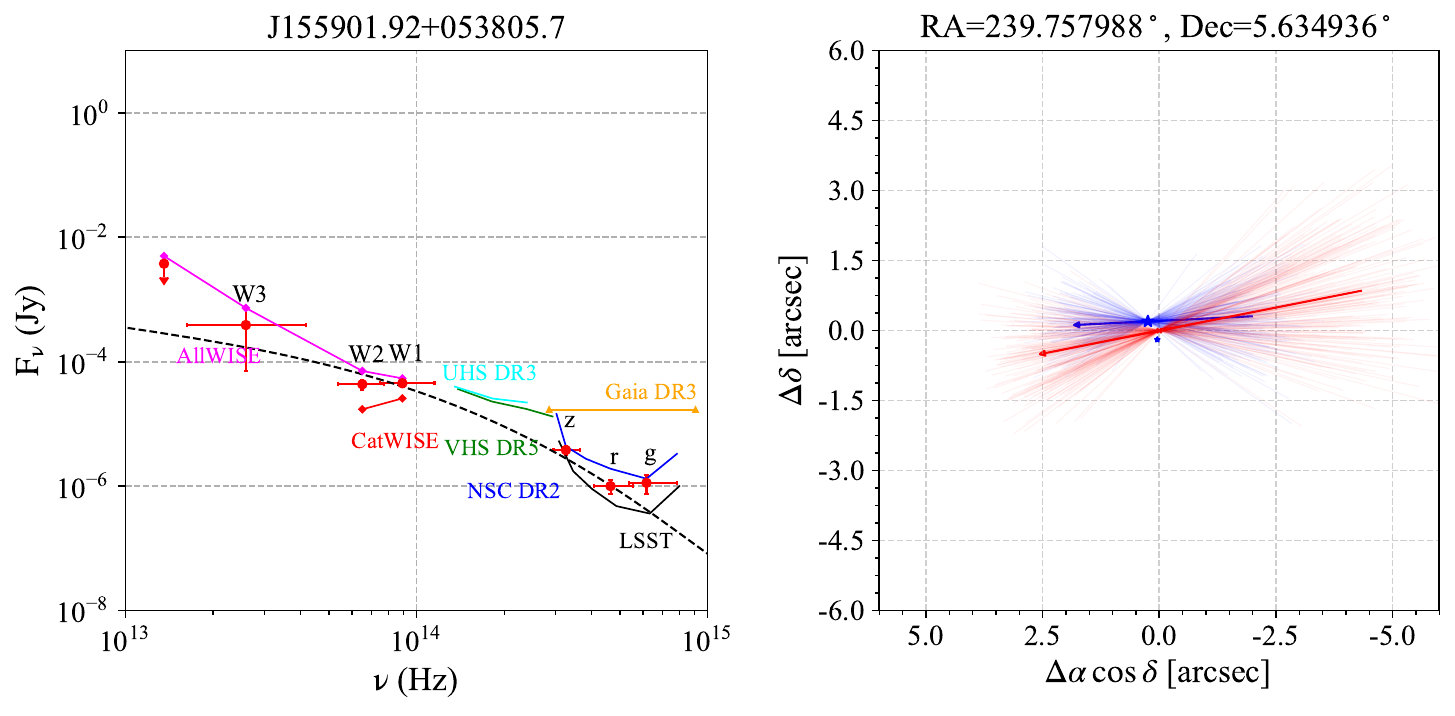}
\caption{Source J155901.92+053805.7.}
\label{f-s8}
\end{figure*}

\begin{figure*}[t]
\centering
\includegraphics[width=0.95\textwidth]{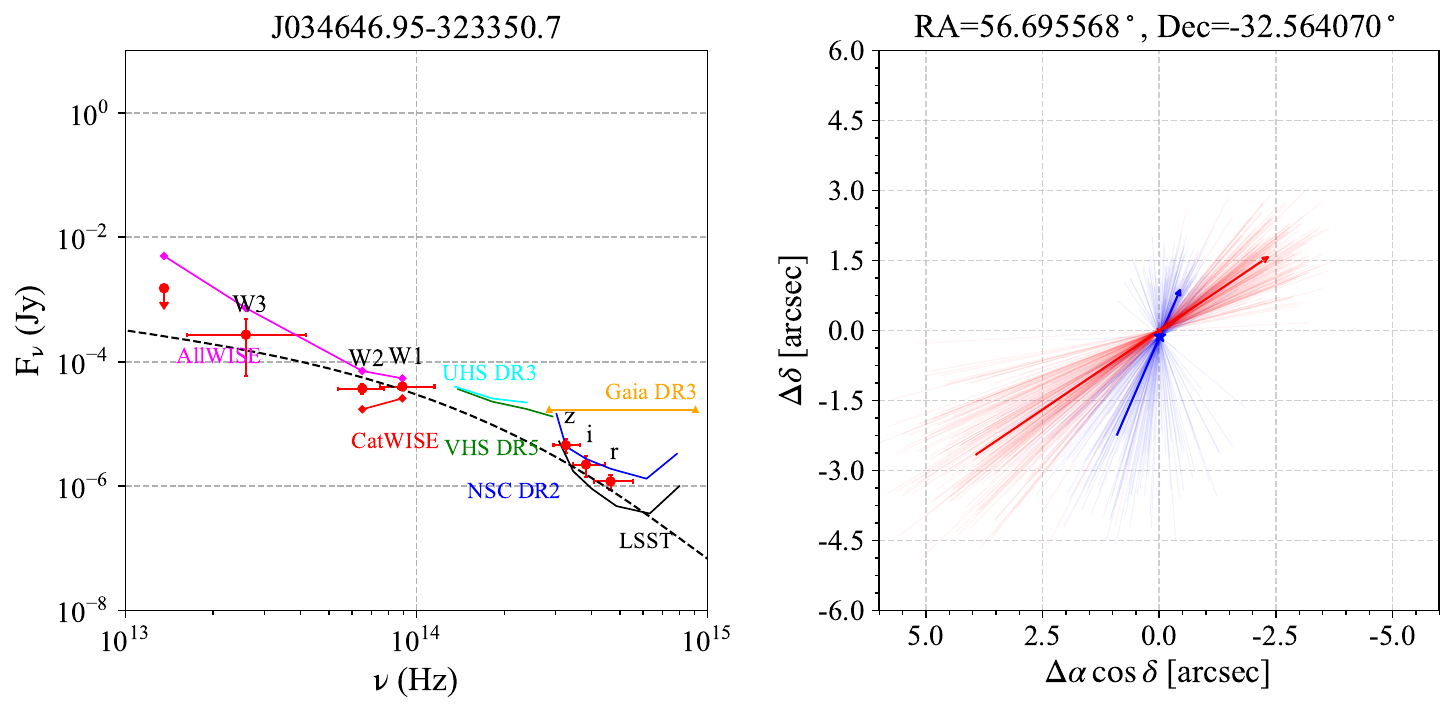}
\caption{Source J034646.95$-$323350.7.}
\label{f-s9}
\end{figure*}

\begin{figure*}[t]
\centering
\includegraphics[width=0.95\textwidth]{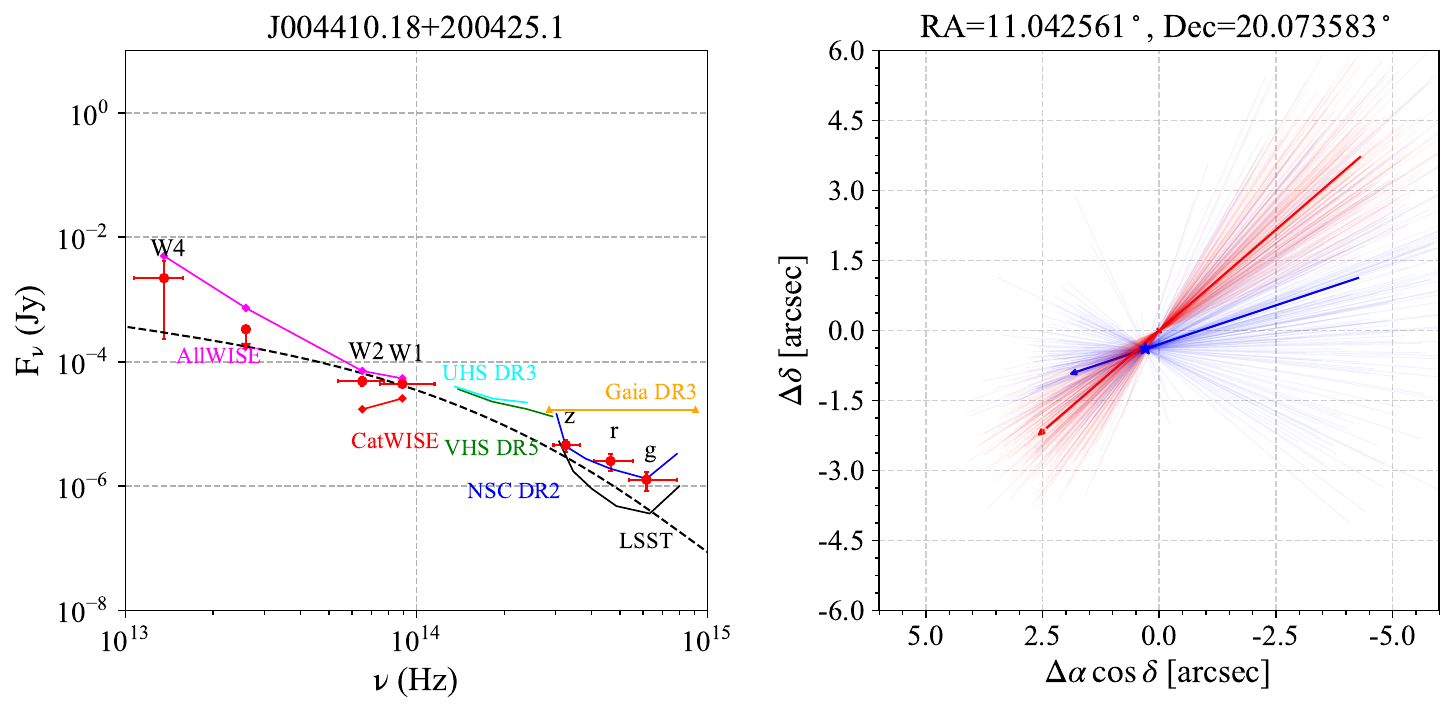}
\caption{Source J004410.18+200425.1.}
\label{f-s10}
\end{figure*}

\begin{figure*}[t]
\centering
\includegraphics[width=0.95\textwidth]{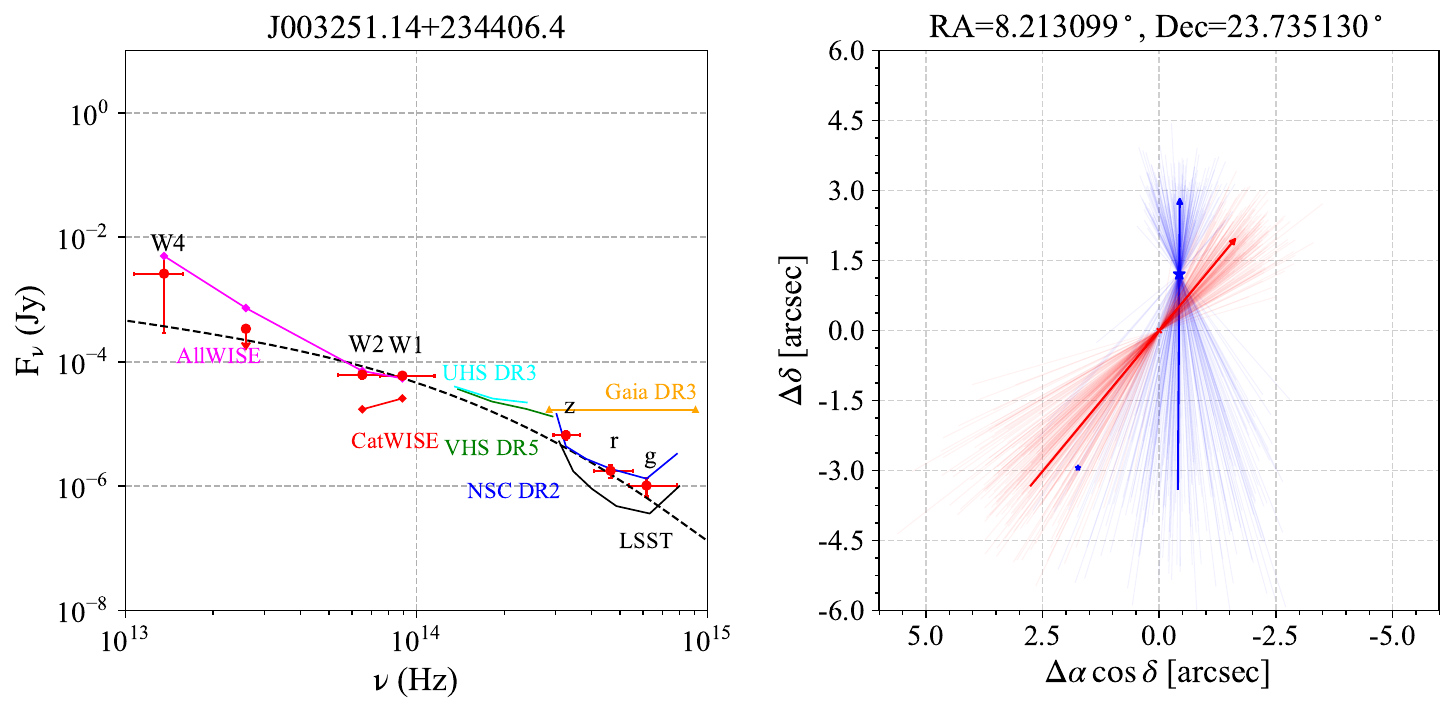}
\caption{Source J003251.14+234406.4.}
\label{f-s11}
\end{figure*}

\begin{figure*}[t]
\centering
\includegraphics[width=0.95\textwidth]{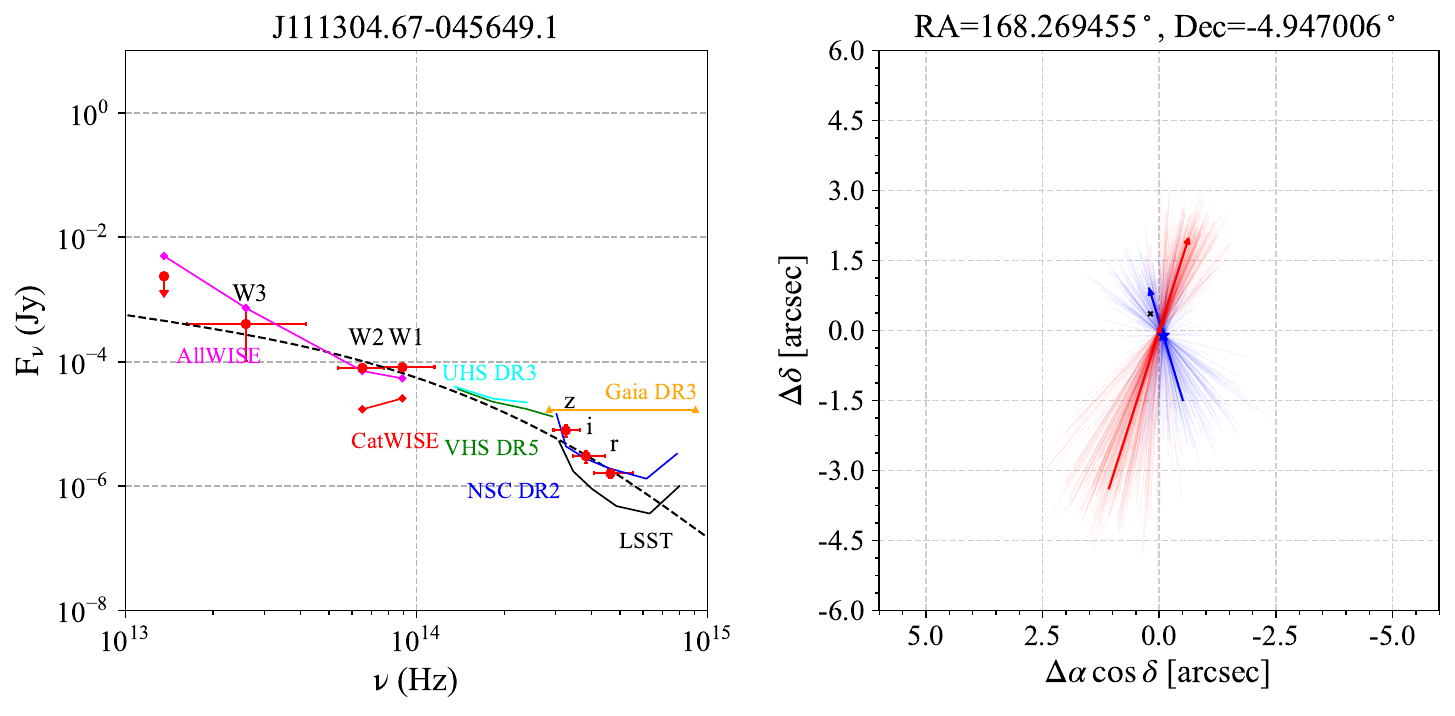}
\caption{Source J111304.67$-$045649.1. The black cross in the right panel indicates a source in VHS~DR5.}
\label{f-s12}
\end{figure*}

\begin{figure*}[t]
\centering
\includegraphics[width=0.95\textwidth]{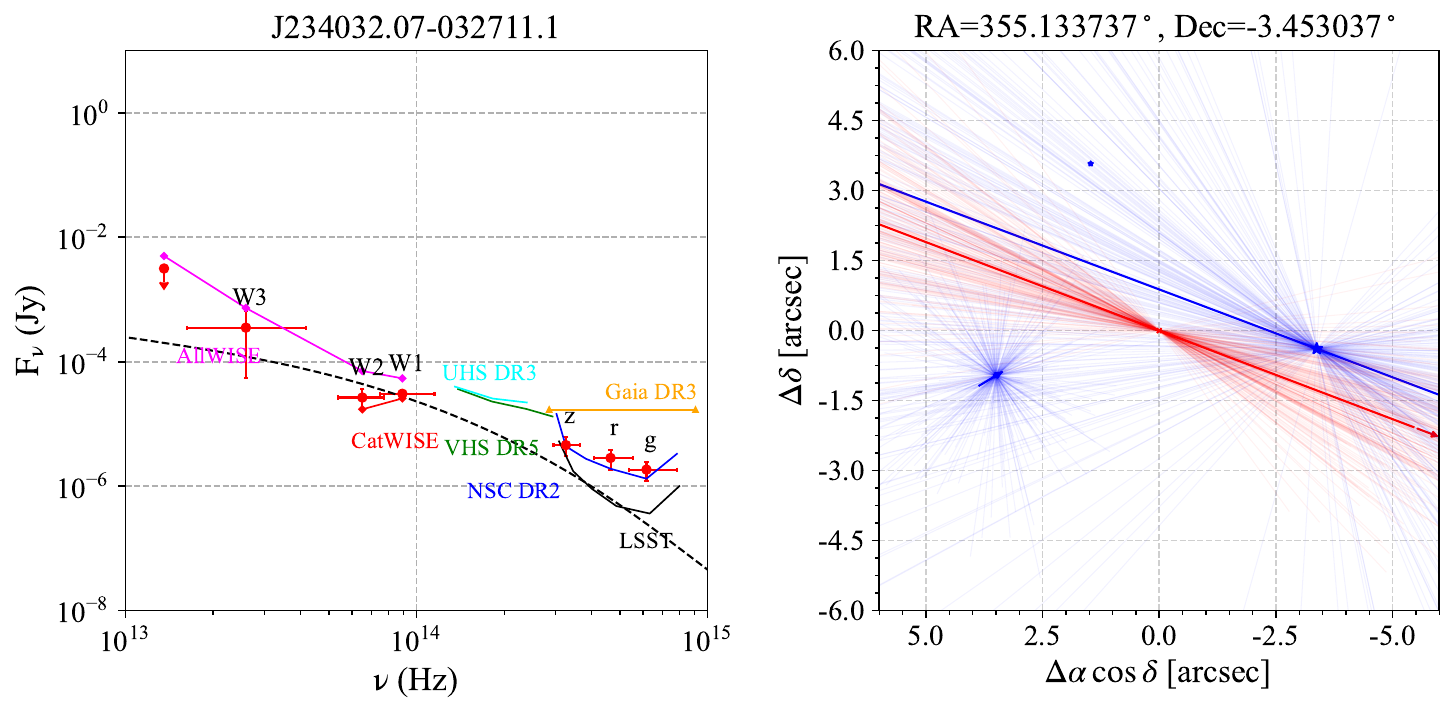}
\caption{Source J234032.07$-$032711.1.}
\label{f-s13}
\end{figure*}

Figs.~\ref{f-s1}-\ref{f-s13} show the gold black hole candidates. Every figure is for a specific candidate. The left panels show the source spectra in the IR and optical bands with flux measurements from AllWISE (W3 and W4), CatWISE2020 (W1 and W2), UHS~DR3, VHS~DR5, and NSC~DR2; there are no flux measurements from UHS~DR3 or VHS~DR5 because none of these sources is in those catalogs. The red dots are the flux measurements; their vertical error bars represent the 2-$\sigma$ flux uncertainties, while their horizontal error bars indicate the band widths. Some vertical error bars are not visible because they are so small that they are within the red dots. Red dots with arrows pointing down indicate upper limits. The black-dashed curves are the best-fits of {\tt LLAGNSED}. The plots also show the nominal detection thresholds of AllWISE (magenta), CatWISE2020 (red), UHS~DR3 (cyan), VHS~DR5 (green), Gaia~DR3 (orange), NSC~DR2 (blue), and LSST (black). The right panels show the maps of the sky regions around our gold candidates. Sources in CatWISE2020 are indicated by red crosses (the positions are those corresponding to the observation time of CatWISE2020), and the red arrows show their predicted trajectories in the sky: the tips of the arrows are at the predicted positions of the sources in 2025 and the opposite ends are at the predicted positions of the sources in 2000. The gold candidates are located at the centers of the plots: $\Delta\delta = \Delta\alpha\cos\delta = 0$. Sources in NSC~DR2 are indicated by blue stars (larger blue stars for the cross-matched candidates and smaller blue stars for other sources, all with their positions at the observation time of NSC~DR2); the blue arrows show the predicted trajectories in the sky: the tips of the arrows are at the predicted positions of the sources in 2025 and the opposite ends are at the predicted positions of the sources in 2000. The thin segments without tips indicate the 1-$\sigma$ uncertainties of these trajectories (red segments for CatWISE2020 and blue segments for NSC~DR2). No sources are reported in Gaia~DR3, VLASS~QL1, RACS-mid, RACS-high, eROSITA~DR2, and SIMBAD in these regions of the sky, while a source is reported in VHS~DR5 near J111304.67$-$045649.1 (see Fig.~\ref{f-s12}, but according to our criteria the VHS source is not associated with our gold candidate).

\clearpage
\newpage
\onecolumngrid

\end{document}